%% file: ms.tex
\documentclass[usenatbib,useAMS]{mn2e}
\usepackage{deluxetable}
\usepackage{epsfig,color}
\usepackage{amsmath}

\newcommand{\gcc}{\ \mathrm{g\ cm^{-3}}}
\newcommand{\cms}{\ \mathrm{cm\ s^{-1}}}
\newcommand{\foe}{\ensuremath{\mathrm{10^{51}\ ergs\ }}}
\newcommand{\nuc}[2]{\ensuremath{\mathrm{^{#1}#2}}}
\newcommand{\ye}{\ensuremath{Y_\mathrm{e}}}
\newcommand{\msun}{\ensuremath{\mathrm{M}_\odot}}
\newcommand{\simgt}{\,\hbox{\lower0.6ex\hbox{$\sim$}\llap{\raise0.6ex\hbox{$>$}}}\,}
\title[SN Ia diversity from central ignition density variations]{Type
  Ia supernova diversity: white dwarf central density as a secondary parameter in three-dimensional delayed detonation models}

\author[Seitenzahl et al. 2010]{I.~R.~Seitenzahl$^{1}$,
  F.~Ciaraldi-Schoolmann$^{1}$, F.~K.~R\"opke$^{1}$\\
$^{1}$Max-Planck-Institut f\"ur Astrophysik,
                 85741 Garching, Germany\\} 

\date{\today}
\pubyear{}
\begin{document}
\maketitle
\begin{abstract}
  Delayed detonations of Chandrasekhar-mass white dwarfs (WDs) have
  been very successful in explaining the spectra, light curves, and
  the width-luminosity relation of spectroscopically normal Type Ia
  supernovae (SNe~Ia).  The ignition of the thermonuclear deflagration
  flame at the end of the convective carbon ``simmering'' phase in the
  core of the WD is still not well understood and much
  about the ignition kernel distribution remains unknown.
  Furthermore, the central density at the time of ignition depends
  on the still uncertain screened carbon fusion reaction rates, the
  accretion history and cooling time of the progenitor, and the
  composition.  We present the results of twelve high-resolution
  three-dimensional delayed detonation SN~Ia explosion simulations
  that employ a new criterion to trigger the deflagration to
  detonation transition (DDT).  The simulations fall into into three
  ignition categories: relatively bright SNe with 5 ignition kernels
  and a weak deflagration phase (three different central densities),
  relatively dim SNe with 1600 ignition kernels and a strong
  deflagration phase (three different central densities) and
  intermediate SNe with 200 ignition kernels (six different central
  densities). All simulations trigger our DDT criterion and the
  resulting delayed detonations unbind the star.  We find a trend of
  increasing iron group element (IGE) production with increasing
  central density for all three categories.  The total \nuc{56}{Ni}
  yield, however, remains more or less constant, even though increased
  electron captures at high density result in a decreasing
  \nuc{56}{Ni} mass fraction of the IGE material.  We attribute this
  to an approximate balance of \nuc{56}{Ni} producing and destroying
  effects.  The deflagrations that were ignited at higher density
  initially have a faster growth rate of subgrid-scale
  turbulence. Hence, the effective flame speed increases faster, which
  triggers the DDT criterion earlier, at a time when the central
  density of the expanded star is higher. This leads to an overall
  increase of IGE production, which off-sets the percental reduction
  of \nuc{56}{Ni} due to neutronization.
\end{abstract}


\section{Introduction}
\label{sec:int}
SNe Ia have come to fame as the Universe's most luminous
standardizable candles -- crucial ingredients to the study of dark
energy and cosmology \citep[e.g.][]{riess1998a,schmidt1998a}.
A limiting factor on the precision of using SNe~Ia as distance
indicators is the inherent scatter in their normalized light curves
\citep[e.g.][]{wood-vasey2007a}.  A better understanding of the
intrinsic variation of supernova brightnesses and spectra is needed 
\citep[e.g.][]{albrecht2006a,miknaitis2007a}. Simulations of SN Ia explosions
are already being used to aid in improving the precision of cosmological distance measurements
based on supernovae in the future \citep[e.g.][]{blondin2011a}. 
In addition, SNe~Ia also play a critical role in
galaxy gas kinematics \citep[e.g.][]{scannapieco2008a}, positron
production \citep[e.g.][]{chan1993a}, and chemical evolution
\citep[e.g.][]{matteucci1986a}.  Detailed modeling of the explosions
is therefore useful for understanding the origin of the Galactic
511 keV line, the origin and evolution of heavy elements, and kinetic
supernova feedback, and measuring the Hubble parameter as a function
of redshift. The standard model of SNe Ia relies on the nuclear fusion
of the initial composition (predominantly \nuc{12}{C} and \nuc{16}{O})
of a massive white dwarf (WD) star to more tightly bound nuclei to
power the explosion \citep{hoyle1960a}.  The exact nature of the
progenitor systems and details of the dynamics of the nuclear burning
processes however are not known. Among the leading scenarios are the
Chandrasekhar-mass models, in which a WD accretes matter from a
companion star and grows in mass to near the Chandrasekhar limit until
pycnonuclear carbon fusion reactions \citep{cameron1959a} start taking
place.  Once carbon fusion reactions produce more energy than is
carried away by neutrino losses, the core becomes convective and when
the nuclear burning time of a fluid element becomes shorter than the
eddy turnover time a deflagration flame may be born
\citep[e.g.][]{woosley1990a}.  Numerical simulations of the convective
stage leading up to the ignition of the deflagration were performed by
\citet{hoeflich2002a}, \citet{kuhlen2006a}, \citet{piro2008b},
\citet{piro2008c}, and \citet{zingale2009a}.  The central density of the WD decreases
significantly during the simmering phase between the onset of carbon
burning and the ignition of the deflagration
\citep[e.g.][]{lesaffre2006a,piro2008c}.  Those calculations, however,
are not taking electron captures and the URCA process correctly into
account, and some uncertainty in the evolution remains.  The rate of
the screened \nuc{12}{C}--\nuc{12}{C} fusion reaction is still quite
uncertain
\citep[e.g.][]{itoh2003a,jiang2007a,gasques2005a,gasques2007a}.  The
central density at the time of ignition, however, depends only mildly
on the exact value of this reaction rate \citep{cooper2009a,
  iapichino2010a}.  More important is the initial mass and the
accretion and cooling history of the WD, which determines the
thermodynamic state of the interior.  This results in a range of
possible central densities at ignition, from less than $2 \times 10^9$
to over $5 \times 10^9 \gcc$ \citep{lesaffre2006a}.

Metallicity has a considerable impact on the supernova brightness
\citep[e.g.][]{timmes2003a,travaglio2005a,bravo2010a}.  In contrast, the ignition
density has been shown to depend rather weakly on metallicity and the
CO ratio \citep{lesaffre2006a}.  If the initial deflagration flame can
transition into a detonation
\citep[e.g.][]{khokhlov1997a,roepke2007d,woosley2007a,woosley2009a},
then good agreement of the models with observations can be obtained
\citep[e.g.][]{roepke2007b,bravo2008a,kasen2009a}.  A successful
explosion model has to reproduce the observed range of peak absolute
magnitudes (i.e. \nuc{56}{Ni} masses) and the width-luminosity
relation and scatter thereabout.  Furthermore, the observed
correlation between the brightness of an event and the delay time or
age of the host stellar population has to be explained
\citep[e.g.][]{gallagher2008a}.  Recently, a connection between age of
the host stellar population and SN~Ia brightness was proposed via the
effect of longer cooling times on the ignition density
\citep{krueger2010a}.  Varying the central density for 150 two
dimensional delayed detonation supernova simulations within the
statistical ignition framework presented in \citet{townsley2009a}, the
authors found that the \nuc{56}{Ni} yield decreased with increasing
central density, while the total iron group element (IGE) yield
remains roughly constant. This is attributed to increased production
of stable isotopes (such as e.g. \nuc{54}{Fe} or \nuc{58}{Ni}) due to
increased neutronization via electron captures at the higher
densities. There are, however, at least three competing effects that
influence the \nuc{56}{Ni} mass produced in a delayed detonation SN.
\begin{enumerate}
\item Electron capture rates on protons and iron-group isotopes under
  electron degenerate condisitons are strongly increasing with density
  \citep[e.g.][]{langanke2001a}.  Consequently, a distribution of
  nuclei in nuclear statistical equilibrium at high density
  neutronizes at a much faster rate than one at lower density
  \citep[e.g.][]{seitenzahl2009a}, which acts to lower the
  \nuc{56}{Ni} mass.
\item Near Chandrasekhar-mass WDs in hydrostatic equilibrium with a
  higher central density are more compact, i.e. significantly smaller
  and slightly more massive and tighly bound. This may translate into
  a more compact WD at the time of the first DDT, which could lead to
  an overall larger part of the WD being burned to IGEs, which acts to
  raise the \nuc{56}{Ni} mass.
\item Deflagrations evolve differently at higher gravitational
  acceleration $g$ \citep{khokhlov1995a,zhang2007a}.  From linear
  stability analysis, the Rayleigh-Taylor temporal growth rate scales
  with $\sqrt{g}$.  The different flame evolution and turbulence
  generation could have an effect on the DDT (e.g. the transition
  density), which, depending on the different degree of
  ``pre-expansion'', could either lower or raise the \nuc{56}{Ni}
  mass.
\end{enumerate}

The effect of variations in the central density of the WD on
\emph{pure deflagrations} has been explored in three-dimensional models before
\citep{roepke2006b}.  Here, we present the results of twelve
high-resolution three-dimensional \emph{delayed detonation} SN~Ia
simulations (that employ a new DDT criterion, see Section
\ref{sec:ddt}) for three different ignition configurations and a range
of central densities.  We find that, for the same spatial ignition
spark distributions, the \nuc{56}{Ni} yield remains more or less
constant a function of central density at ignition.  The
deflagrations that were ignited at higher density produce subgrid-scale turbulence at a higher rate, which triggers the DDT criterion
earlier when the central density of the star is higher. This leads to
an overall increase of IGE production as well as enhanced electron
captures.  Even though the mass in \nuc{56}{Ni} comprises a smaller
fraction of the mass that has burned to IGEs, the overall \nuc{56}{Ni}
yield remains roughly constant since more total mass in IGEs is
produced in the detonation.  Only the cases where much of the IGEs are
produced in the deflagration phase show a trend of decreasing
\nuc{56}{Ni} with central density.  In Section~\ref{sec:sims} we
introduce our setup and briefly review the computational methods, in
Section~\ref{sec:results} and \ref{sec:discussion} we present and
discuss the results, and in Section~\ref{sec:conclusions} we conclude.

\section{Methods and Simulations}
\label{sec:sims}
The large computational demands of the high-resolution three-
dimensional simulations we perform prevented a statistical framework
approach similar to the one presented in \citet{townsley2009a},
\citet{krueger2010a} and \citet{jackson2010a}. Under the constraints
of limited computational resources, we chose six different densities
for a setup with an intermediate number of ignition points (200
kernels), and three densities each for the setups with the least (5
kernels) and the most (1600 kernels) ignition points respectively.
The central densities are such that they cover the distribution of
ignition densities expected from different cooling ages and accretion
histories \citet{lesaffre2006a}.  The ignition spark configurations
are selected in a way that SNe with a range of brightnesses with
\nuc{56}{Ni} masses between $\sim0.45$ and 1.1 \msun\ are obtained.

\subsection{Initial models}
\label{sec:inimod}
All simulations presented here are full star simulations performed in
3D.  The initial stellar models are cold, isothermal ($T=5\times
10^5\,\mathrm{K})$ WDs in hydrostatic equilibrium with central density
$\rho_{\mathrm{c}}$ ranging from $1.0$ to $5.5\times10^9 \gcc$.  The
composition is assumed to be 47.5 per cent \nuc{12}{C}, 50 per cent
\nuc{16}{O}, and 2.5 per cent \nuc{22}{Ne} (to account for solar
metallicity of the zero-age main-sequence progenitor) by mass
homogeneously throughout the star, resulting in an electron fraction
$\ye = 0.49886$.

A strong deflagration phase leads to more energy release and hence
expansion of the star. The ensuing detonation then produces less
\nuc{56}{Ni}, leading to a dimmer event. In the multi-spot ignition
scenarios, the strongest deflagrations are obtained by placing an
optimal number of ignition sparks approximately symmetrically about
the center \citep{garcia2005a,livne2005a,roepke2007c}. While too few
ignition sparks lead to an overall weak deflagration, too many of them
lead to vigorous burning in the initial stage and thus an early
expansion of the WD that suppresses burning in later stages of the
deflagration \citep{roepke2006a}.

Asymmetric ignition spark distributions lead to a weaker deflagration
phase and hence a brighter SN~Ia \citep{roepke2007a,kasen2009a}.
For the ignition of the deflagration we use setups generated from a
Monte-Carlo based algorithm. The primary input parameters are the
number of the ignition kernels and the distribution type.
The details of the ignition process remain unknown. \citet{woosley2004a}
and \citet{wunsch2004a} conclude from analytical models
that multi-spot ignition within the inner $\sim$150~km or so is a possible
scenario. The total number and spatial distribution of the ignition
spots, however, was not conclusively constrained by their models.
We investigate three different sets of explosion models corresponding to
different ignition scenarios.  We choose configurations of 5, 200, and
1600 kernels which are spherically arranged around the center of the
WD following a Gaussian distribution in radius. The placement of
kernels with a distance greater than 2.5 times than a given variance
$\sigma$ is suppressed. For the setups with 5, 200, and 1600 kernels,
we set $\sigma = 0.6$, $0.75$, and $1.8 \times 10^7$ cm,
respectively. The radius of the spherical ignition kernels is set to
$R_{\mathrm{k}} = 10^6$ cm.  Finally, we impose a length scale
$D_{\mathrm{k}}$, which the distances between the centers of the
ignition kernels have to exceed.  $D_{\mathrm{k}}$ is set to $10^6$,
$3 \times 10^5$, and $5 \times 10^4$ cm for the setups with 5, 200,
and 1600 kernels, respectively. Note that for $R_{\mathrm{k}} >
D_{\mathrm{k}}$, the sparks may partially overlap which is the case
for the setups with 200 and 1600 kernels. 
Within a given model suite (i.e. 5, 200, or 1600 kernels),
the locations of the ignition sparks are only once randomly
determined in the beginning; the resulting spatial realization of the ignition
configuration is then kept fixed and identical in all the simulations with
different central density.
With these choices of
ignition spark distributions we cover a large range of \nuc{56}{Ni}
masses, between $\sim0.45$ and 1.1 \msun, which is consistent with
normal SNe~Ia \citep[e.g.][]{contardo2000a,stritzinger2006a,stritzinger2006b}.

\subsection{Computational method}
\label{sec:method}
The reactive Euler equations are solved using a finite volume scheme
based on the PROMETHEUS code by \citet{fryxell1989a}, which is an
implementation of the ``piecewise parabolic method'' (PPM) of
\citet{colella1984a}. The grid resolution is $512 \times 512 \times
512$ cells for all simulations.  We use the expanding hybrid grid
implementation of \citet{roepke2005a,roepke2005b}, with a uniform
inner grid that contains the deflagration level set and a non-uniform
outer grid that covers the remainder of the computational domain. Our
simulation code is based on a large eddy simulation (LES) approach,
which resolves the largest turbulent structures and models the
turbulence on unresolved scales using a turbulence subgrid-scale model
\citep[for details see][]{schmidt2006b,schmidt2006c}. The code uses a
co-moving grid \citep{roepke2005c,roepke2006a} with an outer coarse
grid following the WD's expansion and an inner finer grid tracking the
flame front. The flame itself is treated as a discontinuity separating
fuel and ash; its propagation is tracked with the level set technique
\citep{osher1988a, smiljanovski1997a, reinecke1999a}.  In this thin
flame approximation, the energy liberated in the nuclear burning is
released immediately behind the level set representing the flame
surface. Since nuclear matter burned in a deflagration undergoes
different burning than matter processed in a detonation, separate
level-set representations are used \citep{golombek2005a,roepke2007b}.
Using a full nuclear reaction network in every computional cell to
calculate the source terms for the hydrodynamics is currently still
computationally too expensive for three-dimensional simulations. We
solve this problem by tabulating the energy release as a function of
fuel density. For the detonation we use the new tables from
\citet{fink2010a}. A table for the energy release of the deflagration
level set was calculated in a similar way.

\subsection{DDT criterion}
\label{sec:ddt}
The transition from a subsonic deflagration to a supersonic detonation
based on the Zel'dovich gradient mechanism \citep{zeldovich1970a} was
introduced to SN~Ia theory by \citet{blinnikov1986a} and further
analyzed by \citet{khokhlov1991a, khokhlov1991c, khokhlov1997a} and
\citet{niemeyer1997b}. The main result of their studies was that such
a transition is only possible in the turbulent deflagration stage,
where large velocity fluctuations $v'$ lead to a mixing of cold fuel
and hot ash up to a certain length scale. These ``hot spots" are
supposed to be the seeds of a DDT. \citet{lisewski2000b} pointed out
that $v'$ must exceed $10^8 \cms$. Indeed, velocity fluctuations on
this scale have 
already been found in three-dimensional deflagration simulations \citep{roepke2007d}.
\citet{woosley2007a} argued that for DDTs there are specific
restrictions on the burning properties deep in the distributed burning
regime, which is the regime where strong turbulent flame interactions
are expected. As DDTs cannot be resolved in full-star simulations, we
employ a subgrid-scale model to calculate the probability of these
transitions.

The details of this subgrid-scale DDT model, which is guided
  by the latest studies of the microscopic mechanism of DDTs in SNe~Ia
\cite{woosley2009a}, are described in a
separate paper (Ciaraldi-Schoolmann \& R\"opke, in preparation). It
accounts for the intensity of the turbulent velocity fluctuations as
well as the fuel density $\rho_{\mathrm{fuel}}$ and fuel fraction
$X_{\mathrm{fuel}}$ in the grid cells crossed by the flame front.
If
the probability $P (v' > v'_{\mathrm{crit}})$ to find velocity
fluctuations larger then $v'_{\mathrm{crit}}$ in a specific area
$A_{\mathrm{flame}}$ at the flame front exceeds a certain threshold
$A_{\mathrm{crit}}$, detonations are ignited in the grid cells which
contain the largest velocity fluctuations. $A_{\mathrm{flame}}$ is
defined as the part of the flame where $\rho_{\mathrm{fuel}} =
[0.6-0.8]\cdot 10^7 \gcc$ and $X_{\mathrm{fuel}} = [0.3 - 0.7]$. To
properly estimate this area, we take the fractal dimension of the
flame front into account, which is $\sim 2.36$
\citep{sreenivasan1991a, kerstein1988a, woosley2007a}. The number of
ignitions is given by the ratio $A_{\mathrm{flame}}$ to
$A_{\mathrm{crit}}$.  The criterion must hold at least for half of an
eddy turn $\tau_{\mathrm{eddy}_{1/2}} = L/V(L)$, where $L$ is the
turbulent integral scale and $V(L)$ the velocity at this
scale. Following considerations of \citet{woosley2007a},
\citet{ciaraldi2009a} and \citet{roepke2007d} we assume $L = 10^6$ cm
and $V(L)\approx [10^7 - 10^8] \cms$. We choose a constant value of
$\tau_{\mathrm{eddy}_{1/2}} = 0.005$ s in our analysis.  We further
follow \citet{lisewski2000b} and \citet{roepke2007d} and define
$v'_{\mathrm{crit}} = 10^8 \cms$ and $A_{\mathrm{crit}} = 10^{12}
\mathrm{cm}^2$ as our thresholds for the DDT-criterion. While
  the details of the implementation are beyond the scope of this
  publication and will be presented elsewhere, we point out that this modeling approach is
  significantly different from simply fixing a certain DDT threshold
  density. Our criterion in addition requires strong local turbulent
  velocities. It is thus more restrictive and substantially reduces the number of
  DDTs.

\begin{figure}
  \includegraphics[width=\columnwidth,clip]{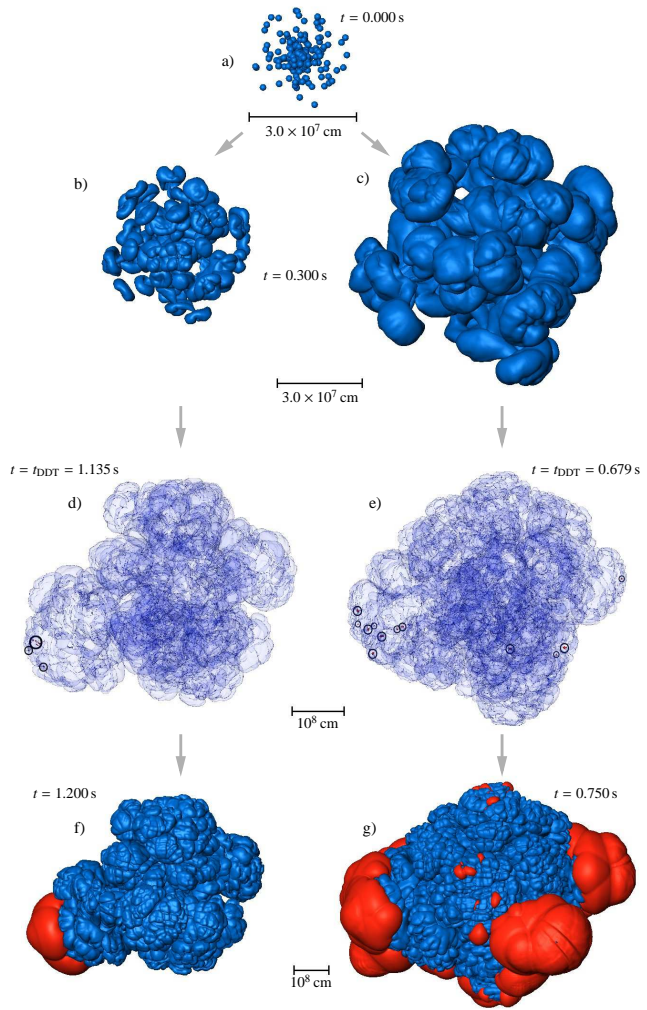}
  \caption{Shown are snapshots of the deflagration level set (blue)
    and the detonation level set (red) for the lowest (left column,
    $\rho_{\mathrm{c}}=1.0 \times 10^9 \gcc$) and the highest (right
    column, $\rho_{\mathrm{c}}=5.5 \times 10^9 \gcc$) central
    density. a) Both simulations initially have a spatially identical
    arrangement of the 200 ignition kernels. b/c) After
    $t=0.3\mathrm{s}$, the deflagration has burned significantly more
    for the high density case. d/e) The spots where the DDT criterion
    is first triggered are circled; the detonation triggers at an
    earlier time for the high density case. f/g) The detonation level
    set is propagating through unburned fuel away from the DDT
    spots. Multiple detonations can be launched as long as the DDT
    criterion is fulfilled. In spite of the differences in time
    evolution and morphology, both models produce the same amount of
    \nuc{56}{Ni}.}
  \label{fig1}
\end{figure}

\section{Results}
\label{sec:results}
\begin{figure}
  \includegraphics[angle=270,width=\columnwidth,clip]{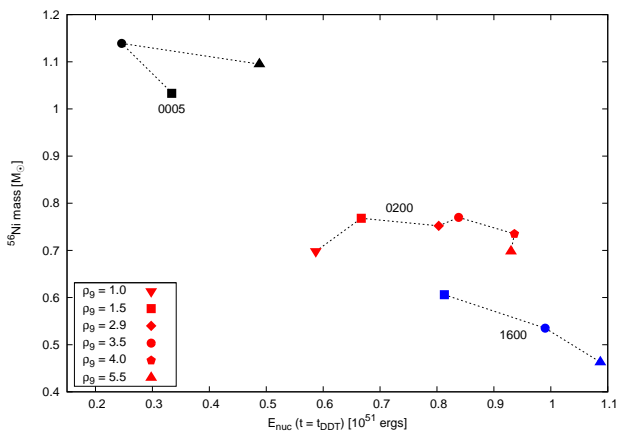}
  \caption{Shown is the mass of \nuc{56}{Ni} produced in the different
    explosions as a function of the total nuclear energy liberated
    during the deflagration phase up to $t=t_{\mathrm{DDT}}$.}
  \label{fig2}
\end{figure}
\begin{figure}
  \includegraphics[angle=270,width=\columnwidth,clip]{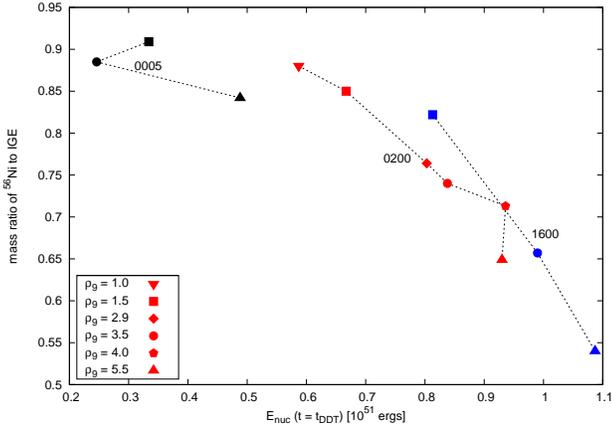}
  \caption{Shown is the relative mass fraction of \nuc{56}{Ni} to IGE
    produced in the different explosions as a function of the total
    nuclear energy liberated during the deflagration phase up to
    $t=t_{\mathrm{DDT}}$. It is evident that the strength of the
    deflagration (as measured by
    $E_{\mathrm{nuc}}(t=t_{\mathrm{DDT}})$) is a very good proxy for
    the mass ratio of \nuc{56}{Ni} to IGE.}
  \label{fig3}
\end{figure}
\begin{figure}
  \includegraphics[angle=270,width=\columnwidth,clip]{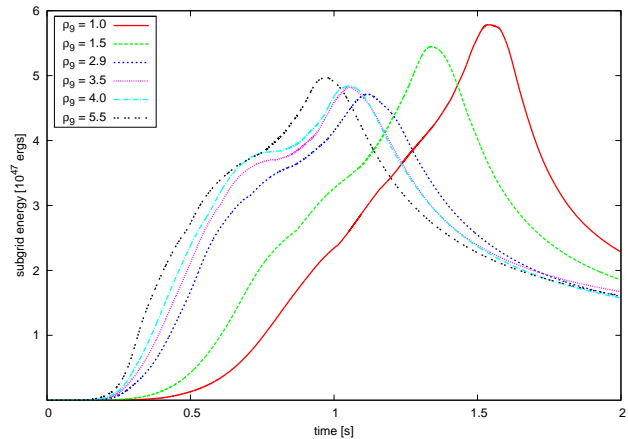}
  \caption{Shown is the subgrid-scale turbulent energy as a function
    of time for the model sequence with 200 ignition kernels. Note
    that the rate of turbulent energy production is initially larger
    for the high density cases.}
  \label{fig4}
\end{figure}
\begin{figure}
  \includegraphics[angle=270,width=\columnwidth,clip]{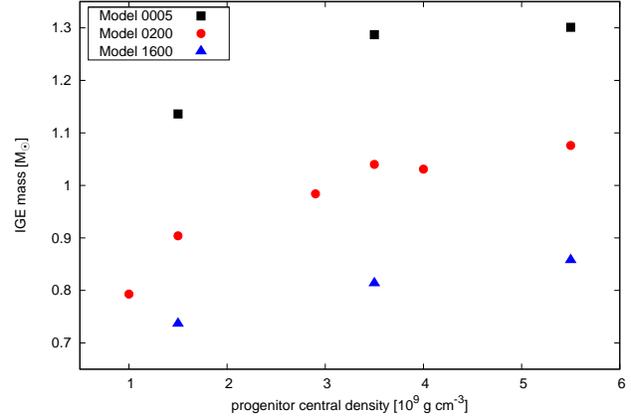}
  \caption{Shown is the mass of IGE as a function of central density
    $\rho_{\mathrm{c}}$ of the WD at the time the deflagration was
    ignited. A trend with increasing IGE mass with central density is
    evident for all three ignition configurations.}
  \label{fig5}
\end{figure}
\begin{figure}
  \includegraphics[angle=270,width=\columnwidth,clip]{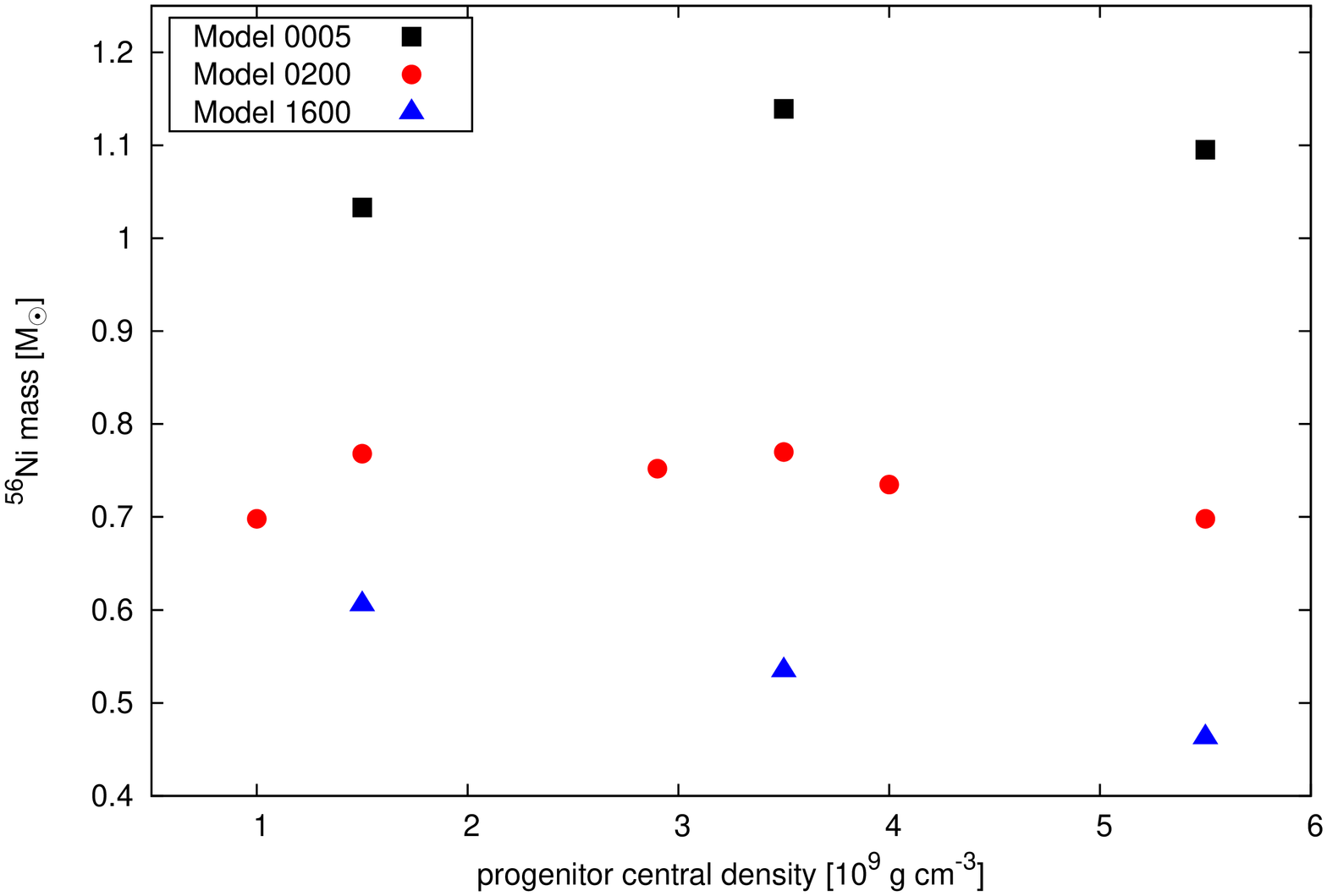}
  \caption{Shown is the mass of \nuc{56}{Ni} as a function of
    $\rho_{\mathrm{c}}(t=0)$. Evidently, for these simulations the
    ignition configuration is the primary parameter that determines
    the \nuc{56}{Ni} mass (and hence peak brightness) of the events,
    whereas the central density is merely a secondary parameter.}
  \label{fig6}
\end{figure}

The chosen distinct setups lead to different evolutions of the
deflagration flame.  In turn, the different evolutions of the flame
front has great impact of the properties of the following delayed
detonations (see Fig.~\ref{fig1}).

On the one hand, the energy released during the deflagration stage up
to the time $t_{\mathrm{DDT}}$ when the first DDT occurs, is smaller
for simulations starting with fewer ignition kernels (see
Table~\ref{tab1}).  The ensuing weaker expansion manifests itself in a
higher central density at $t=t_{\mathrm{DDT}}$.  This in turn
translates into a larger fraction of the total mass of the WD burned
into IGE during the detonation phase (see Table~\ref{tab2}) than for
the models with more ignition kernels.  We thereby confirm that the
strength of the deflagration is a primary parameter for the
\nuc{56}{Ni} production \citep{roepke2007b}, and hence brightness, of
a supernova that explodes in the delayed detonation scenario (see
Fig.~\ref{fig2}). A strong deflagration results in fainter events,
primarily due to the large expansion of the star prior to the DDT, but
also due to the copious neutronization in the deflagration phase (see
Fig.~\ref{fig3}).

On the other hand, for an identical spatial distribution of ignition
kernels, the simulations with higher initial central density exhibit
stronger turbulence production and the subgrid-scale energy grows at a
faster rate initially (see Fig.~\ref{fig4}). This can be understood by
the difference in gravitational acceleration $g$. For the same
\emph{spatial} distribution of ignition spots, the distribution in the
\emph{mass} coordinate will be centered further out at larger mass and
thus larger $g$ in the higher central density case.  For
Rayleigh-Taylor dominated deflagration flames the turbulent burning
velocity scales with $\sqrt{g}$ \citep[e.g.][]{khokhlov1995a}.
Consequently, at equal time after ignition (e.g. $0.3 \mathrm{s}$), a
high central density model will have burned significantly more mass
than a low central density model (compare Figs.~\ref{fig1}b and
\ref{fig1}c).  As a further consequence, the higher the initial
central density, the sooner our DDT criterion is fulfilled (see
Table~\ref{tab1}, but note the outlier with 5 ignition kernels for
$\rho_{\mathrm{c}}=3.5\times10^9 \gcc$). Importantly, there is a trend
that models with high initial central density also have a higher
central density at $t=t_{\mathrm{DDT}}$, which is a proxy for the
amount of fuel at densities high enough that it will be burned to IGE
in the detonation.  \input{tab1.tex} \input{tab2.tex} We
find that for all ignition kernel distributions, the total yield of
IGE material within an ignition distribution suite increases with
$\rho_{\mathrm{c}}$ (see Fig.~\ref{fig5} and Table~\ref{tab2}).  The
total yield of $\nuc{56}{Ni}$ appears flat with $\rho_{\mathrm{c}}$
for the model suites with 5 and 200 ignition kernels; only the model
suite with the strongest deflagration phase (1600 ignition kernels)
has a trend of decreasing \nuc{56}{Ni} with $\rho_{\mathrm{c}}$ (see
Fig.~\ref{fig6} and Table~\ref{tab2}). 
%Notably, these trends differ
%from the results of \citet{krueger2010a}, who find that the yield of
%IGE material remains constant with $\rho_{\mathrm{c}}$ and that the
%\nuc{56}{Ni} yield is decreasing with $\rho_{\mathrm{c}}$.

\section{Discussion}
\label{sec:discussion}
We attribute the almost constant \nuc{56}{Ni} mass to a coincidental
balance of the competing effects presented in Section~\ref{sec:int},
-- the larger electron capture rates at higher central density are
offset by a greater total mass in IGE due to the inherent compactness
of the WD and faster evolution of the flame towards DDT.  The high
density simulations with 5 and 200 ignition sparks exhibit more
subgrid-scale energy generation at early times and therefore higher
flame speeds initially. The still highly turbulent deflagration flame
reaches the outer layers of the WD with low fuel density faster, and,
consequently, the DDT criterion triggers earlier when the central
density of the star is still higher (see Table~\ref{tab1}). As a
result, these simulations produce more IGE and even though the
\nuc{56}{Ni} fraction of the IGEs is lower due to increased
neutronization, (see Fig.~\ref{fig3}), the total amount of
\nuc{56}{Ni} remains roughly constant.  The difference between
\nuc{56}{Ni} and IGE is mainly made up of stable iron group nuclides
such as \nuc{58}{Ni}, \nuc{54}{Fe}, but also other radioactive
nuclides such as \nuc{57}{Ni} or \nuc{55}{Co} contribute.

The model suite with 1600 ignition kernels, which has such a high
density of ignition sparks that the whole central region is filled
with burning products of the deflagration, behaves differently.  Due
to the numerous ignition sites, a large part of central mass of the
star is already burned in the deflagration before the first DDT
occurs.  The IGE produced in the deflagration, where most of the
electron captures occur, are a large fraction of the total IGE
produced (see Table~\ref{tab2}).  The electron captures and resulting
shift of \nuc{56}{Ni} toward more neutron rich stable Fe-group
isotopes occur most copiously behind the slowly moving deflagration
flame front, and consequently the increased production of IGE material
does not reflect in larger \nuc{56}{Ni} masses for cases where the
deflagration contributes most of the IGE mass.  The strong
deflagration and vigorous expansion leads to such low central
densities at $t=t_{\mathrm{DDT}}$ that the ensuing detonation cannot
produce sufficient \nuc{56}{Ni} to counter this trend. Events
producing such large amounts of strongly neutronized IGE matter
cannot, however, make up most SN~Ia events, due to the unusual
isotopic composition \citep[e.g.][]{woosley1997a}.

We can only speculate why our simulations predict increasing IGE and
roughly constant \nuc{56}{Ni} production for higher initial central
density, whereas \citet{krueger2010a} find the opposite -- decreasing
\nuc{56}{Ni} and constant IGE.  One possible reason for the different
trends between the two sets of simulations is the way the DDT is
handled.  For simulations of delayed detonations in SNe occurring via
a DDT, the detonation is generally put in ``by hand''
\citep[e.g.][]{arnett1994b,livne1999a,gamezo2005a,bravo2008a,krueger2010a}
, usually by choosing a critical density where a deflagration
transitions to a detonation.  Recently, \citet{jackson2010a}
investigated the effect the particular choice of such a transition
density has.  They found a quadratic dependence of the IGE yield on
the log of the transition density.  Instead of imposing a fixed
transition density, we utilize a dynamic DDT criterion (see
Section~\ref{sec:ddt}), which takes the effects of different
deflagration evolutions on the detonation initiation into account. We
point out that typical densities where our DDT criterion triggers (see
Table~\ref{tab1}) are lower than $10^7 \gcc$.  \citet{jackson2010a}
have shown that the variance of the \nuc{56}{Ni} yield for a
statistical set of simulations is relatively large for such a low
choice of transition density (see figure~3 from their work), in
agreement with our observed large range of \nuc{56}{Ni} masses
obtained.  Numerous other obvious differences between the simulation
sets exist, including the nature of the propagation and the nuclear
energy release of the burning fronts (level sets vs. reaction progress
variables), the dimensionality of the simulations (3D vs 2D), or the
structure of the computational mesh (AMR vs. expanding grid).

In this context, note the work of \citet{meakin2009a}, who present a
suite of supernova explosion models with different off-sets for the
initial deflagration bubble. Although their single bubble off center
ignition scenario does not explore central density at the time of the
ignition of the deflagration as a parameter, their result that a
strong deflagration phase need not necessarily result in less
\nuc{56}{Ni} produced is the same. They also find that the total
amount of IGE decreases for models that had a more vigorous
deflagration phase (leading to more expansion) before the detonation
is triggered, but the \nuc{56}{Ni} yields remains approximately
constant (see figure 12 of their work).

\section{Conclusions}
\label{sec:conclusions}
We have performed twelve three-dimensional hydrodynamical simulations
for delayed detonation SNe~Ia for a range of central densities and
ignition conditions. We find a trend of increasing IGE
production with central density within each set of ignition
conditions.  This is because the high central density WDs are more
compact and the flame evolves faster; the DDT occurs sooner when more
unburned material is still above the density threshold ($\approx 10^7
\gcc$) where a detonation will still produce IGE.  In spite of the
larger IGE mass, the more vigorous neutronization occurring in the
high density models during the deflagration phase yields \nuc{56}{Ni}
massest hat are more or less constant with $\rho_{\mathrm{c}}$ for the
brighter SNe. Only dim SNe, which have a strong deflagration phase and
expansion prior to the DDT, exhibit a trend of decreasing \nuc{56}{Ni}
mass with increasing density, since the increased neutronization in
the deflagration phase cannot be compensated for by the relatively
weak detonation phase. This trend, however, is of secondary importance
when compared to the effects of varying the ignition kernel
distribution.  For a given ignition kernel spatial distribution, the
central density therefore influences the brightness of the supernova
event only as a secondary parameter. From the works of
\citet{townsley2009a} and \citet{bravo2010a}, it appears that the same
holds for composition, i.e. metallicity and C/O ratio. Indeed, based
on an analysis of high-quality V and B-band light curves of SNe~Ia
from the Carnegie Supernova Project, \citet{hoeflich2010a} propose
that the composition and central density are two independent secondary
parameters for SN~Ia light curves.  In light of the importance of the
ignition configuration of the deflagration for the brightness of the
SN, it is most crucial to establish how the central density at
ignition (cooling time) and metallicity affect the statistical
properties (notably number and location) of the ignition sparks
themselves, and not their respective direct effects on the outcome of
an explosion once a random ignition spark distribution was chosen.
One should therefore aim to quantify which effect composition, cooling
and accretion history have on the ignition process, for example by
mapping them into the exponentiation parameter $C_{\mathrm{e}}$ of the
stochastic ignition prescription of \citet{schmidt2006a}.  This would
require a better understanding of the physics leading up to ignition,
including the nature of the convection and effects of electron
captures and the convective URCA process.

\section*{Acknowledgements}
The simulations presented here were carried out as part of the DEISA
grant ``SN-DET-hires'' on facilities of the Leibnitz Rechenzentrum, in
part on the JUGENE supercomputer at the Forschungszentrum J{\"u}lich
within project HMU13, and in part at the Computer Center of the Max
Planck Society, Garching, Germany.  This work was supported by the
Deutsche Forschungsgemeinschaft via the Transregional Collaborative
Research Center TRR 33 ``The Dark Universe'', the Emmy Noether Program
(RO 3676/1-1), and the Excellence Cluster EXC~153. We also want to
thank Wolfgang Hillebrandt for reading the manuscript and his helpful
comments. 

\bibliography{../../bibliography/bib/astrofritz_MNRAS}
\bibliographystyle{../../bibliography/bst/mn2e}
%\bibliography{../../../bibliography/bib/astrofritz}

\end{document}

%% file: tab1.tex
\begin{table*}
\caption{DDT attributes for all models. Tabulated are the nuclear
    energy released $E_{\mathrm{nuc}}$, the central density $\rho_{\mathrm{c}}$, and  the
    (average) density of the first DDT spot(s) $\bar{\rho}_{1}(t=t_{\mathrm{DDT}})$
    at the time $t_{DDT}$ when the first DDT(s) occured. \label{tab1}}
\begin{tabular}{cccccc} \hline
{Model}&$\rho_{\mathrm{c}}(t=0)$&$t_{\mathrm{DDT}}$&$E_{\mathrm{nuc}}$($t=t_{\mathrm{DDT}}$)&$\rho_{\mathrm{c}}(t = t_{\mathrm{DDT}})$&$\bar{\rho}_1(t=t_{\mathrm{DDT}})$\\ 
       &$[10^9 \gcc]$&[s]&[\foe]&$[10^8 \gcc]$&$[10^7 \gcc]$\\ \hline 
        0005 &1.5&1.253&0.334&4.911&0.773\\ 
         ... &3.5&0.890&0.246&11.610&0.758\\
         ... &5.5&0.911&0.488&6.183&0.716\\ \hline
        0200 &1.0&1.135&0.587&1.450&0.700\\
         ... &1.5&0.993&0.667&1.706&0.746\\
         ... &2.9&0.802&0.803&2.270&0.756\\
         ... &3.5&0.756&0.838&2.533&0.749\\
         ... &4.0&0.755&0.936&2.193&0.759\\
         ... &5.5&0.679&0.930&2.711&0.761\\ \hline
        1600 &1.5&1.077&0.813&0.718&0.705\\
         ... &3.5&0.848&0.990&0.827&0.779\\
         ... &5.5&0.757&1.087&0.875&0.755\\ \hline
\end{tabular}
\end{table*}

%% file: tab2.tex
\begin{table*}
\caption{Nucleosynthetic yields for all models. Tabulated are the
  total WD mass $M_{\mathrm{tot}}$, and the final masses of \nuc{12}{C}, \nuc{16}{O}, intermediate mass
  elements, iron group elements, and \nuc{56}{Ni}
  ($M_{\nuc{12}{C}}$,$M_{\nuc{16}{O}}$,$M_{\mathrm{IME}}$,$M_{\mathrm{IGE}}$,
  and $M_{\nuc{56}{Ni}}$). Furthermore tabulated are the masses of iron
  group elements and \nuc{56}{Ni} at the time $t_{\mathrm{DDT}}$
  when the first DDT(s) occured ($M_{IGE}^{\mathrm{def}}$ and $M_{\nuc{56}{Ni}}^{\mathrm{def}}$), as well as their respective relative
  fractions of the final masses,
  ($\tfrac{M_{\mathrm{IGE}}^{\mathrm{def}}}{M_{\mathrm{IGE}}}$ and $\tfrac{M_{\nuc{56}{Ni}}^{\mathrm{def}}}{M_{\nuc{56}{Ni}}}$). \label{tab2}}
\begin{tabular}{cccccccccccccc} \hline
{Model}&$\rho_{\mathrm{c}}(t=0)$&$M_{\mathrm{tot}}$&$M_{\nuc{12}{C}}$&$M_{\nuc{16}{O}}$&$M_{\mathrm{IME}}$&$M_{\mathrm{IGE}}$&$M_{\nuc{56}{Ni}}$&
$M_{\mathrm{IGE}}^{\mathrm{def}}$&$M_{\nuc{56}{Ni}}^{\mathrm{def}}$&$\tfrac{M_{\mathrm{IGE}}^{\mathrm{def}}}{M_{\mathrm{IGE}}}$&$\tfrac{M_{\nuc{56}{Ni}}^{\mathrm{def}}}{M_{\nuc{56}{Ni}}}$\\ 
       &$[10^9 \gcc]$&[\msun]&[\msun]&[\msun]&[\msun]&[\msun]&[\msun]&[\msun]&[\msun]&&\\ \hline 
        0005  &1.5&1.378&0.004&0.033&0.205&1.136&1.033&0.221&0.182&0.19&0.176\\
         ...  &3.5&1.406&0.002&0.017&0.100&1.287&1.139&0.168&0.125&0.13&0.110\\
         ...  &5.5&1.416&0.002&0.015&0.098&1.301&1.095&0.288&0.172&0.22&0.157\\ \hline
        0200  &1.0&1.361&0.017&0.104&0.447&0.793&0.698&0.385&0.321&0.49&0.460\\
         ...  &1.5&1.378&0.012&0.073&0.390&0.904&0.768&0.447&0.348&0.49&0.453\\
         ...  &2.9&1.400&0.008&0.063&0.346&0.984&0.752&0.547&0.359&0.56&0.477\\
         ...  &3.5&1.406&0.006&0.053&0.307&1.040&0.770&0.573&0.354&0.55&0.460\\
         ...  &4.0&1.409&0.007&0.057&0.314&1.031&0.735&0.622&0.375&0.60&0.510\\
         ...  &5.5&1.416&0.007&0.053&0.280&1.076&0.698&0.626&0.325&0.58&0.466\\ \hline
        1600  &1.5&1.378&0.014&0.104&0.523&0.737&0.606&0.478&0.366&0.65&0.604\\
         ...  &3.5&1.406&0.015&0.098&0.479&0.814&0.535&0.600&0.346&0.74&0.647\\
         ...  &5.5&1.416&0.015&0.094&0.449&0.858&0.463&0.665&0.310&0.78&0.670\\ \hline
\end{tabular}
\end{table*}